# Multiple Benefits through Smart Home Energy Management Solutions—A Simulation-Based Case Study of a Single-Family-House in Algeria and Germany

**Marc Ringel [1,\*], Roufaida Laidi [2,3] and Djamel Djenouri [2]**

[1] Energy Economics, Nuertingen Geislingen University, 73312, Geislingen, Germany;

[2] CERIST Research Center, Ben Aknoun, Algiers 16028, Algeria; ddjenouri@acm.org (D .D.); rlaidi@cerist.dz (R. L.)

[3] Ecole nationale Supérieure d'Informatique (ESI), Oued-Smar, Algiers 16309, Algeria

\* Correspondence: marc.ringel@hfwu.de

**Abstract:** From both global and local perspectives, there are strong reasons to promote energy efficiency. These reasons have prompted leaders in the European Union (EU) and countries of the Middle East and North Africa (MENA) to adopt policies to move their citizenry toward more efficient energy consumption. Energy efficiency policy is typically framed at the national, or transnational level. Policy makers then aim to incentivize microeconomic actors to align their decisions with macroeconomic policy. We suggest another path towards greater energy efficiency: Highlighting individual benefits at microeconomic level. By simulating lighting, heating and cooling operations in a model single-family home equipped with modest automation, we show that individual actors can be led to pursue energy efficiency out of enlightened self-interest. We apply simple-to-use, easily, scalable impact indicators that can be made available to homeowners and serve as intrinsic economic, environmental and social motivators for pursuing energy efficiency. The indicators reveal tangible homeowner benefits realizable under both the market-based pricing structure for energy in Germany and the state-subsidized pricing structure in Algeria. Benefits accrue under both the continental climate regime of Germany and the Mediterranean regime of Algeria, notably in the case that cooling energy needs are considered. Our findings show that smart home technology provides an attractive path for advancing energy efficiency goals. The indicators we assemble can help policy makers both to promote tangible benefits of energy efficiency to individual homeowners, and to identify those investments of public funds that best support individual pursuit of national and transnational energy goals.

**Keywords:** energy efficiency; smart building; multiple benefits; smart environments; homeowner benefits

## 1. Introduction

Energy efficiency has become established as a key pillar of energy policy in many countries for the simple reason that energy saved—sometimes coined 'negawatts' or 'negajoules'—contributes to supply security, competitive energy prices and environmental protection [1]. This has led the International Energy Agency (IEA) to use the term "first fuel" to describe energy efficiency, considering it to be a major energy resource, but one still largely untapped. In 2009, their long-term analysis predicted that by 2035 a full two-thirds of the economic potential of energy savings would be lost unless policy activity was increased [2].

The benefits of energy savings are compelling. Whether focusing on energy security, sustainable growth, climate change, or the well-being of the citizenry, policy debates recognize energy efficiency as a key goal [3–6]. This is especially true in the European Union (EU) and countries of the Middle East and North Africa (MENA) [7]. The EU recently adopted a revision of their energy policy in which it introduced the "energy efficiency first principle". This means that before making any "energy planning, policy and investment decisions", the decision-makers should consider whether "alternative energy efficiency measures could replace in whole or in part the envisaged planning, policy and investment measures" [8]. The German government has adopted an analogous approach to its long-term strategies [9–11], relying on research that supports the macroeconomic value of thinking "energy efficiency first" [12–14].

Still, the microeconomic level is where human actors make decisions to implement energy efficiency upgrades. But the potential benefits of efficiency are realized only after going through the pain of pursuing its implementation: the confusion over the many upgrade options, the need for research or expert assistance, the time invested, the inconvenience of construction, and of course the financial outlay. Microeconomic actors face high hurdles to aligning their actions with macroeconomic policy goals. This may explain why a recent analysis by the European Commission found that three out of every four buildings in Europe are energy inefficient [15,16]. The situation is similar in the MENA countries.

While bridging this efficiency gap requires actions by owners of both commercial and residential properties, it is the single-family homeowner whose decisions, considered collectively, will have major consequences for the energy future. This is the conclusion that the governments of both Germany and Algeria came up with a joint energy efficiency plan in a bilateral agreement for the exchange of best practices and addressing common energy challenges [17]. The untapped "negawatt" potential represents great savings for both countries [18], where the single-family home represents the majority of housing stock. In Germany, 15.6M out of 18.8M residential buildings (82.9%) are single-family-homes, representing 2.2b $m^2$ of living area [19]. In Algeria, 62.1% of the housing stock is single-family homes [20]. These account for 40% and 44% of the total final consumption of energy in Germany and Algeria, respectively [17,21,22].

The single family home is also where the greatest potential is found for energy upgrades. In Germany, 11.72M (64%) of the homes were built before 1978, the year in which energy performance standards for new buildings were introduced [23–26]. On average, these buildings consume 208 kWh/$m^2$ energy per annum [24,21,10] and represent the country's greatest potential for saving energy and reducing greenhouse gas (GHG) emissions [27] .

Moreover, work by the IEA [1], the United Nations Environment Programme (UNEP) [28] and the US Environmental Protection Agency (EPA) [29] has shown that energy efficiency has benefits beyond "negawatts" and GHG mitigation. Energy efficiency can produce a range of benefits from improving public health to creating new jobs, leading to what the IEA calls the "multiple benefits approach" to quantifying policy impact. Such analyses establish even more strongly the value of energy efficiency [29–31].

These analyses, however, rest on macroeconomic arguments. Recent attempts to translate the benefits to the microeconomic level of the household (e.g., [32]) have remained theoretical. Still, these studies argue rightly that benefits need to be established at the level of the individual dwelling where investment decisions are made. What remains unknown, however, is how macroeconomic benefits from energy efficiency become microeconomic benefits to the homeowner. Which of the multiple macroeconomic benefits would best translate into tangible homeowner rewards?

This question is no longer as theoretical as it once was. The emergence of smart home technology has opened the way to efficiency upgrades that do not require major financial commitments. Today's smart home technology permits a lower-cost approach to energy efficiency—namely, making more efficient use of existing equipment in the home. This is possible even without the grand promises of Internet-of-Things (IoT) technology. Wireless architectures including these technologies use real time information collected by sensors and processed by AI algorithms in the cloud to provide real time control that preserves energy and satisfy occupant's comfort. In addition, the IoT technology enables the user to monitor and control his house with his mobile devices and remotely over internet. Furthermore, it is not unreasonable to think that a homeowner who sees tangible benefits from using smart home technology might be motivated to pursue more costly efficiency upgrades, perhaps replacing an old furnace or adding new insulation. Hence, the smart home path can also be seen as a stepping-stone to realizing broader policy goals.

The aim of our study is to determine which of the multiple macroeconomic benefits of energy efficiency could be realized by and promoted to a single-family homeowner willing to make a modest investment in smart home technology. We developed a set of benefit indicators that align well with standard key performance indicators (KPIs) used in commercial energy assessments. We did so by simulating one year of lighting, heating and cooling operations in a single-family home under a baseline scenario and two smart home scenarios. Our model situated the homes in the regulatory environments of Algeria and Germany. This allowed us to check our benefit indicators for robustness and practicality under different regulatory conditions. We think our results can contribute to policy design by identifying key benefits available to homeowners considering investments in energy efficiency.

The rest of this paper is structured as follows: Section two of this paper reviews research on the multiple benefits of energy efficiency in the building sector; our intent is to place those benefits in the context of the individual homeowner. On that basis, we develop our methodology, described in section three, with its guide for assessing economic, environmental and social benefits associated with smart home efficiency initiatives. Section four presents the results of our simulations, which we discuss in section five. In Section six, we present our conclusions and recommendations.

## 2. Multiple Benefits of Energy Efficiency

The literature on the economic impact of energy either not used or not produced dates back to 1990 [33], following the energy-savings debates brought on by the oil crises of the 70's. By the early 2000's, the concept of economic savings through energy efficiency had entered the climate policy discussion as "ancillary benefits" or "co-benefits" to climate mitigation action [34–36]. In the years since, the multiple benefits of energy efficiency have been more widely recognized by policy analysts, bolstered by the research and advocacy efforts of the International Energy Agency (IEA) and the United Nations Environment Programme (UNEP). IEA's "multiple benefits approach" is similar to many evaluation schemes, e.g., the Impact Assessment Guidelines of the European Commission [37], and similar regulatory approaches in many countries, including Algeria and Germany [30,38–40].

### 2.1. Multiple Benefits in Energy Policy

Figure 1 presents on the left an overview of the original IEA concept of multiple benefits; on the right is the concept adapted to a framework of green growth [41–43] by Ürge-Vorsatz et al. following the notion of "multiple impacts" or MI [44].

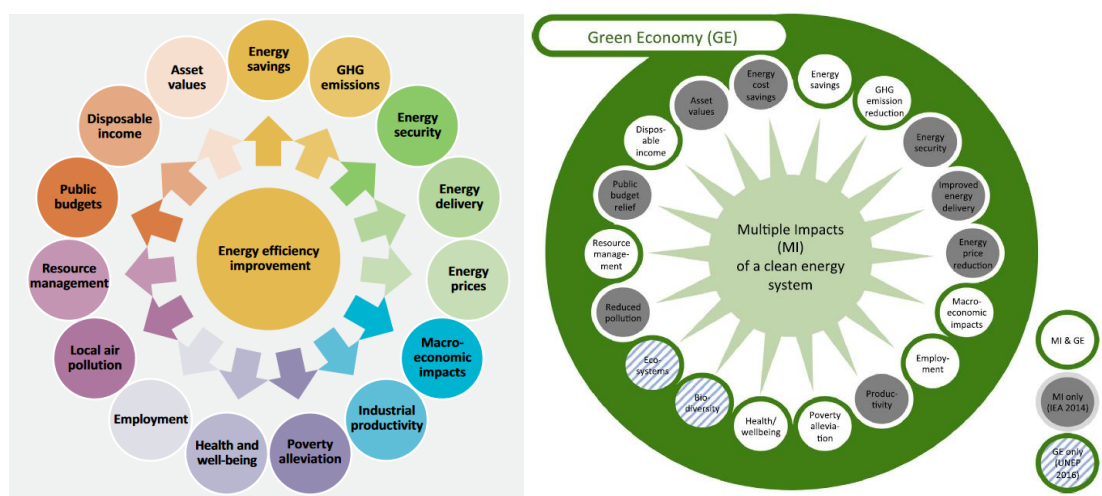

**Figure 1.** Concepts of multiple benefits: IEA concept and relation to green growth. Sources: [1,44].

Evident here is the broad reach of efficiency benefits. As Kerr et al. [7] note, this has allowed policymakers in different countries to more comprehensively quantify the benefits associated with energy efficiency, and so more fairly evaluate the cost-benefits trade-off that drives many policy decisions [12,13].

Most recently, the EU research framework programme Horizon 2020 funded the COMBI project (Calculating and Operationalising the Multiple Benefits of Energy Efficiency in Europe) to quantify the multiple non-energy benefits of energy efficiency in Europe. The project quantified the impact of efficiency out to the year 2030 by modelling 21 sets of "energy efficiency improvement" (EEI) actions; in total, 31 individual impact indicators were quantified, more than half of which can be monetized [45,46]. Table 1 summarizes these by category.

**Table 1.** Multiple impacts of additional EU energy efficiency policies (annual; 2030 time horizon).

| Category | Benefit | Identified Benefits |
|---|---|---|
| Resources | Energy savings | 257 TWh avoided power generation from combustibles |
| | Energy security | Improved energy security of up to 5%; reduced import costs of fossil fuel in the amount of 48bn € |
| | Other resource savings | 850 Mt of material resources saved |
| Economic | Investment | 11bn € avoided investment in generation plants |
| | GDP | 1% rise in GDP (additional 161bn € GDP) |
| | Jobs | Creation of 2.3m job-years |
| | Decrease in fuel prices | 1.3% oil; 2% coal; 2.9% natural gas |
| | Public budgets | Additional €86m available for other policies |
| | Productivity of the economy | 39m additional workdays (4.7bn €) |
| Social | Premature deaths due to indoor cold | 3000—24,000 avoided |
| | Avoided disability-adjusted life years due to indoor dampness and related asthma | 2700—22,300 |
| Environmental | Avoided premature deaths due to PM 2.5 particulates | >10,000 |
| | Avoided premature deaths due to O3 particulates | 442 |
| | Avoided direct CO2eq emissions | 300 Mt |

Source: Authors' own compilation, based on [46].

The benefit valuations show significant spread, because no standardized valuation approach exists for all the impacts [47]. Different categories, even different benefits require different valuation methods. Although each valuation may be input into a cost-benefit analysis (CBA), it can be difficult to ensure a uniform quality of inputs. Ürge-Vorsatz et al. [44] list 60 different methodologies that have been used to assess physical, monetary and non-monetary impacts of improved energy efficiency. Moreover, monetized impacts cannot simply be summed together in a CBA, because they may possibly overlap or interact [48,49].

At the national or regional level, macroeconomic tools such as partial or general equilibrium models, input-output-analysis or econometric models are used to assess the impacts of energy efficiency policies. These models can be combined in hybrid approaches to cover impacts that otherwise would be difficult to quantify [29,50,37]. Such efforts allow public authorities to more comprehensively balance the benefits of energy efficiency initiatives against the public costs incurred. As valuable as this may be to the policy analyst working from a macroeconomic perspective, it does not communicate multiple benefits to the homeowners who make the decision to invest in energy efficiency upgrades.

## 2.2. Multiple Benefits for Homeowners

Assessing the multiple monetary and non-monetary benefits of energy efficiency for the homeowner requires analysing the situational factors that influence individual energy consumption, including geography, weather patterns, energy performance of the dwelling, occupational patterns and comfort demands. The complexity of such analysis has prompted a range of studies into homeowner benefits. These are summarized in Table 2 by benefit category.

**Table 2.** Homeowner benefits of increased energy efficiency.

| Category | Benefit | Studies Available (Selection) |
|---|---|---|
| Resources | Energy savings | [51–53] |
| | Freshwater savings | [54,55] |
| | Other resource savings ("grey energy") | [56] |
| Economic | Energy cost savings (including taxes) | [57,58] |
| | Lower operation costs | |
| | Access to government subsidies | [59] |
| | Increased asset value | [60] |
| | Cost-effective investment decision | [61] |
| Social | Improved indoor air quality | [62,63] |
| | Increased amenity or convenience, higher comfort levels | [64,65] |
| | Health benefits | [66–68] |
| | Safety/burglary prevention/monitoring | [69] |
| | Higher disposable income; reduction of individual energy poverty | [70,71] |
| Environmental | $CO_2$-reductions | [72] |
| | Reduced local air pollutants | [73,74] |
| | Reduced noise levels | [75] |
| | Waste and wastewater reduction | |

Source: Authors' own compilation based on [32,46,76,77].

These studies typically analyse specific impacts, rather than using a case study approach to assess impacts comprehensively across the four benefit categories. The results provide insight into potential benefits, but lack the specificity needed to offer a homeowner concrete facts on which to base decisions. This stems not only from the limited datasets, but also from the lack of a standardized set of tools and techniques for determining the value of energy efficiency benefits to a homeowner.

Existing home energy audits illustrate the challenge. An energy audit gives a homeowner a breakdown on the sources of energy consumption in the home; the auditor can then offer advice, identify energy improvement options, and even provide refurbishing roadmaps [78,79] to guide upgrade investments. Homeowners also have online calculators available to estimate levels of energy consumption and related greenhouse gas (GHG) emissions before-and-after a proposed upgrade [80,25]. These permit a partial assessment of individual benefits and contributions to GHG mitigation, but like the energy auditors they focus on monetized energy savings. The multiple individual benefits associated with increased energy efficiency—such as burglary prevention or automated health monitoring or increased occupant comfort—are ignored. Yet these are often key to guiding investment decisions, notably in home automation [69]. Their assessment through a traditional energy audit would require precise measurements and verification. Even if this were feasible, the results would be limited to the individual case.

However, with software that has emerged to serve the energy industry, it is possible to simulate energy consumption in a home at the detailed level of daily operations. A prototypical single-family home can be modelled, allowing for the assessment of the multiple benefits available to a homeowner willing to pursue energy efficiency through home automation. The multiple benefits derived are analogous to those that would be gained were the homeowner to pursue traditional efficiency upgrades.

The simulation approach allows us to derive a set of easy-to-use indicators suitable for use in a CBA that has been extended to take into account the external benefits complementing the monetary benefit of reduced energy costs. By considering the energy efficiency profile for a single-family-home that turns "smart" by installing smart lighting, heating and cooling systems, we demonstrate that based on standard profiles, multiple benefits can be derived that remain realizable even under diverse climatic conditions.

## 3. Methods

We investigated the multiple benefits of energy efficiency available to a homeowner by simulating the use of smart home technology to manage energy consumption in a single-family home. We built simulation models using commercially available software and modelled home automation based on easily obtainable smart home components. We situated our simulated cases in Germany and Algiers, using commercially available software to model the two divergent climates. Indicators for the totality of multiple benefits available to the homeowner were then derived.

### 3.1. Simulation Study

We chose to simulate a single-family home for our analyses. We did so for two reasons. First, the single-family home represents the majority of housing stock in Germany and Algeria, as was discussed in the introduction. Second, it offers the greatest potential for energy upgrades, especially in Germany where the bulk of the housing stock was constructed before energy efficiency regulations went into effect in the construction industry.

The single-family home we simulated has a total area for heating and cooling of 150 m² over two levels and follows a two-bedroom, one-bath house plan. This plan was provided by a template included in the 3-D Modeller of the *DesignBuilder* [81] software package we used in our assessments. We linked the model output to the energy simulation program *EnergyPlus* [82], which provided the weather inputs used in the simulations, inputs based on historical meteorological data from the two regions.

The simulation scenarios were as follows:

- *Baseline*: Lighting is controlled manually by family members who turn off the lights during vacation periods. A traditional thermostat with a fixed set point controls heating and cooling.
- *Low-cost*: Lighting is controlled by a smart lamp following a set family schedule for at-home and away periods. The same schedule is used by a smart thermostat to control temperature set points, but the smart thermostat also optimizes settings by learning household patterns over time.
- *Extended*: Lighting is controlled by motion sensors that detect human presence. Daylight harvesting is provided, modelled through sensors that interface with the smart lamp. Further control of the heating and cooling system is provided by an auto-away feature that detects long periods of non-occupancy. Also simulated is the response of the user to suggestions sent by the smart thermostat to adjust the set point for further energy savings.

In all the simulation scenarios, the dwelling was assumed to house a family of four. The occupancy pattern of the family included regular non-occupancy during the week representing work hours and random occupancy during the day on weekends. We also simulated one month of holidays where the family leaves the house 15 days in the winter and 15 days in the summer. This roughly mirrors holiday patterns in Algeria and Germany.

Our simulation also included a factor ignored in most other studies, namely the increased demand for cooling made by the household in response to the increasingly common hotter days. We modelled this as an increase in cooling energy consumption of the dwelling when the interior temperature exceeds 25 °C. Including the household's demand for greater cooling in this way makes a significant impact on the overall energy consumption, revealing even more strongly the tangible benefits gained from better management of home energy consumption.

## 3.2. Commercial Smart Devices

This section provides a description of the devices modelled in the scenarios of our simulation study. The features described are offered by devices available in 2019; we selected for simulation those features related to energy management. The devices may be categorized into smart light bulbs, smart thermostats, context-aware sensors, and wireless hubs:

- *Smart light bulbs*: These devices are light bulbs that are internet connected. Some connect via Wi-Fi, meaning a homeowner needs no extra hardware. Others need a hub connected to a router. They offer lighting control with considerable flexibility. Schedules can be set off daily schedules, and outputs can be set to different levels based on preferences or the input from light sensors. The latter allows for harvesting existing day light in lieu of burning electricity.
- *Smart thermostats*: These are similar to programmable thermostats that allow users to program their preferred temperature settings over a 7-day schedule. Smart thermostats, however, have learning capabilities. As such, simulations can consider the dynamic adjustment of settings in response to weather conditions and occupancy patterns. Smart thermostats may also include sensors that detect long periods of vacancies and allow for adjusting the heating and cooling in response. These devices also may include smart notifications to inform the user about the possibility of saving energy by changing the current settings, a feature we modelled under the extended scenarios.
- *Context-aware sensors*: We integrated motion and light sensors in the simulation scenarios, the former being the most commonly used type. These are designed to detect the presence or absence of people in a room, and so offer finer scheduling granularity than predefined calendar and day settings. Light sensors allow lighting systems to respond to available light conditions.
- *Wireless Hubs*: Commercially available smart devices communicate through different protocols, so wireless hubs are used to integrate these heterogeneous devices under centralized control.

In configuring the smart devices, we assumed that a single hub for lighting as well as a central thermostat controlling both heating/cooling were enough to control the whole house, with one control item needed per room for the other devices. The investment cost of each installation depends on the additional hardware required by the scenario. Table 3 shows these costs, the devices they include, and the home operation simulated. In the baseline scenario, all necessary equipment is in place, so no costs arise. In the low-cost scenario, a scheduling system operates thermostats and lighting according to a pre-defined schedule. In the extended scenario, sensor data drives the heating and cooling equipment, as well as the lamps. Costs listed represent the lowest prices found for the smart devices, based on the manufacturers' websites and two major online marketplace platforms.

**Table 3.** Simulated Scenarios: Costs and Operation.

| Scenario | Investment Cost (€) | Smart Devices | Home Operation |
|----------|---------------------|---------------|----------------|
| Baseline | -/- | -/- | Manual control |
| Low-cost | 268.93 | Thermostat, lamp | Fixed schedule control |
| Extended | 528.35 | Thermostat, lamp, hub, motion & light sensors | Sensor-based control |

## 3.3. Homeowner Benefit Indicators

In the following, we derive our indicators for the multiple benefits gained through smart management of energy consumption in the home. We align our indicators with existing legislative frameworks, specifically the methods and performance indicators presented in ISO 52016 (the earlier ISO 13790) [83,84] and the European Commission's Delegated Regulation No. 244/2012 [85], proposed to assess the cost optimality of building efficiency when implementing the EU Directive for the Energy Performance of Buildings [83,86].

### 3.3.1. Resource Consumption Indicators

A key indicator of energy efficiency improvement is the calculation of energy savings, $\Delta E_a$, over a one-year period. The annual energy savings is derived using Equation (1):

$$\Delta E_a = E_{ref} - E_i \tag{1}$$

where, $E_{ref}$ is the annual energy consumption of the reference scenario (baseline situation); and $E_i$ is the energy consumption of solution $i$.

This implies that if the $\Delta E$ value is positive, the solution, $i$, consumes that much less energy over one year than the reference scenario. To derive total avoided energy consumption, this indicator can be made into a lifecycle assessment by cumulating the energy savings over the lifetime ($\Delta E_T$) of the measure, as shown in Equation (2):

$$\Delta E_T = \sum_{t=1}^{T} (E_{ref} - E_i), \tag{2}$$

where, $t$ is the number of years in the lifecycle assessment (10 years in our study).

We chose a 10-year-lifespan for analysis to allow comparisons to related works that span roughly the same period (usually between eight and 15 years) [87]. The indicator of (2) represents a simple-to-use version of a full life cycle assessment (LCA) [88,89], and we use it to derive not only the lifetime savings benefit to the homeowner but also the contribution made by efficiency gains to protection of the environment. The indicator does not, however, consider mitigated effects to energy savings such as rebound effects [90]. The first reason is that rebound effects are still hard to quantify and depend largely on the individual and cultural context, which in our study is quite diverse; second, we expect building automation and smart home control to largely address behavioural rebound effects.

### 3.3.2. Economic Indicators

To assess the economic performance of the smart home solutions, we apply Cost-Benefit-Analysis (CBA) [91,92], which is essentially the approach taken to appraise building performance by both the ISO and the EU "cost optimal refurbishing" methodology [83,85,93]. The elements of a standard CBA provide several indicators that capture the economic rationale behind a decision to adopt or reject a given home solution. The *payback period (PB)* is the first of these, defined (Equation (3)) as the time to recoup the investment in a smart home installation:

$$PB \text{ [number of years]} = \frac{\Delta In_i}{\Delta OC_i} \tag{3}$$

where, $\Delta In_i$ are the additional costs for installation solution, $I$, over the reference scenario; and $\Delta OC_i$ are the averaged energy savings related to the installation of solution, $I$ (it should be noted that the definition of the PB indicator follows the framework established by the European Commission in its Impact Assessment and related background studies for the energy performance of buildings Directive [93,94]. An alternative definition is the period $t$ in which the NPV computed is zero).

The payback period is favoured for its simplicity, or as a point of reference for budgeting. In principle, the shorter the payback period, the more likely homeowners will implement the solution. However, as Araújo et al. [95] point out, payback period is not a sufficient basis on which to compare solutions. The time value of money is not considered; alternative options might be available that represent better investments. Hence, to better capture the homeowner's budget decision, we consider the net *present value (NPV)* of the investment, as well as the *internal rate of return (IRR),* given in Equations (4) and (5) respectively:

$$NPV = \sum_{t=1}^{T} \left( \frac{\Delta OC_t}{(1+r)^t} \right) - \Delta In_i, \tag{4}$$

$$IRR = -\Delta In_i + \sum_{t=1}^{T} \left( \frac{\Delta OC_t}{(1+r)^t} \right) \stackrel{\text{def}}{=} 0, \tag{5}$$

where $r$ is the discount rate applied to operating costs.

Equation (4) defines NPV as the difference between the present value of energy cost savings over the analysis period and the initial investment required to realize those savings. It represents the monetary value today to the homeowner of the investment. A positive NPV means the expected savings generated by the smart home investment exceeds the cost of the investment. While homeowners can steer clear of smart home installations with a negative NPV, and further can compare two installations on the basis of their NPV, they need more than this to make an investment decision, namely the IRR value (%) calculated by setting NPV equal to zero. The homeowner can use this as an economic rate of return to compare against other available investments.

In our simulations, we used a discount rate of 5% to reflect price increases, as well as other barriers against placing the money into the smart home automation (transaction costs). The rate was fixed following the practice in the literature (e.g., [95,87]) to make results comparable. In order to derive operating costs, energy prices of € 0.3048 per kWh for electricity (used for lighting and cooling) and € 0.0609 per kWh for natural gas (used for heating) were used, in line with the present costs of energy in Germany [21]. For the Algerian case, we followed the pricing policy defined by the government where the first 125 KWh of consumed electricity costs € 0.014 per kWh, and € 0.033 per kWh for all remaining consumption. Similarly, the first 1125 TWh of natural gas are charged € 0.0012 per unit, and € 0.0024 after that [96]. The low prices of energy in Algeria stem from subsidies provided by the government.

### 3.3.3. Social Indicator

Table 2 in Section 2 lists a number of social indicators, but most of them are context-specific, such as indoor air quality or higher comfort levels. As home automation becomes more widespread and datasets from indoor sensors emerge, statistics may develop that serve as reasonable indicators. But for this analysis, we chose to focus on a single social indicator: additional disposable income (ADI) for energy poor households. The indicator is derived based on the principle that upfront investment will be provided by social benefits or government subsidies. It follows that the ADI is represented directly by the effective inflow resulting from lower annual operation costs for the smart home solution in comparison to the reference scenario (Equation (6)):

$$ADI = \sum_{t=1}^{T} \left( \frac{\Delta OC_t}{(1+r)^t} \right). \tag{6}$$

Clearly, this indicator is a less direct proxy for social benefit than the indicators for the previous two benefit categories. Still, it can be argued that focusing on the additional disposable income (or in this case, the cumulated monetary savings) can strengthen the case to energy poor households for engaging in home automation. It becomes an even stronger indicator in favour of automation-based efficiency improvements when comfort effects, i.e., additional cooling demands, are added to the home's electricity consumption.

### 3.3.4. Environmental Indicators

The macroeconomic benefits of efficiency on the environment typically preclude individual homeowners from tracing contributions to the mitigation of negative environmental impacts directly to their use of energy efficient equipment [97,98]. For this analysis, we derived environmental impacts at the homeowner level by applying environmental conversion factors established by the German Environment Protection Agency [99] to the annual energy consumption (Table 4). This allowed us to derive environmental impacts for $CO_2$ emissions, methane, fine dust and particulate matter as well as for local air, water and soil pollutants.

**Table 4.** Specific emission factors of energy use.

| Emission Type | Unit | Emission Coefficient (2016) |
|---|---|---|
| Sulfur dioxide | g/kWh | 0.290 |
| Nitrogen dioxide | g/kWh | 0.440 |
| Particulate matter | g/kWh | 0.017 |
| $PM_{10}$ | g/kWh | 0.015 |
| Carbon monoxide | g/kWh | 0.230 |
| $CO_2$ | kg/kWh | 0.516 |
| NO | g/kWh | 0.013 |
| $CH_4$ (methane) | g/kWh | 0.184 |
| Volatile organic compounds (VOCs) | g/kWh | 0.017 |
| Mercury | mg/kWh | 0.010 |

Source: [99].

Again, the total environmental benefit per emission type ($\Delta EB_e$) accrued by applying a given smart home solution over the reference case can be derived based on the aggregated energy saving (Equation (7)):

$$\Delta EB_e = ec\left[\sum_{t=1}^{T}(E_{ref} - E_i)\right], \tag{7}$$

where, $ec$ is the emission coefficient of the specific emission type from Table 4.

Refining the indicator would be needed to adapt the emission coefficients to real annual values. We expect those rates to be established by public authorities as big data continues to accumulate [100–102]. For our analysis, the indicator can serve as a good proxy for "minimum individual environmental impact" made by the homeowner's efficiency initiatives.

## 4. Results

We developed indicators to assess the energy savings, investment return, social impact and environmental impact of energy efficiency upgrades via home automation. The following presents the resulting calculation of these indicators over 10-years of simulated operation of a single-family home in Algiers, Algeria and Stuttgart, Germany, where two different climate regimes and two different regulatory environments exist.

### 4.1. Energy Saving Impacts of Applying Home Automation Equipment

Table 5 as well as Figures 2 and 3 present the results related to annual energy consumption.

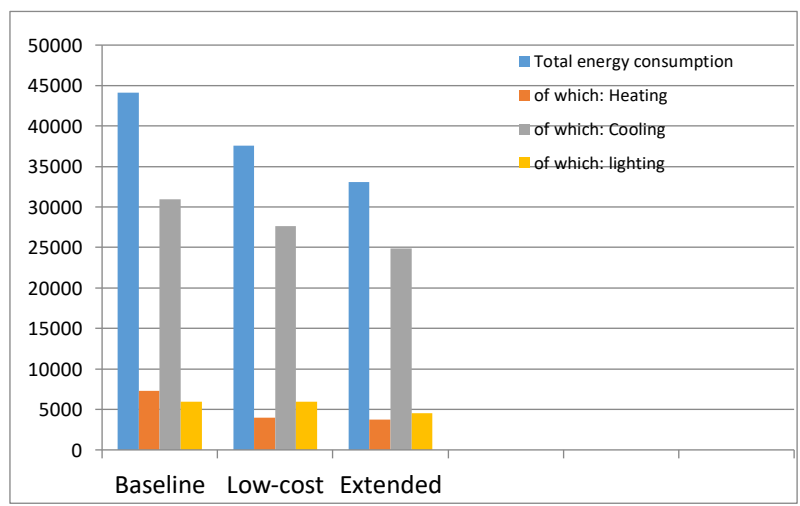

**Figure 2.** Annual energy consumption for heating, cooling and lighting; case of Algiers (kWh).

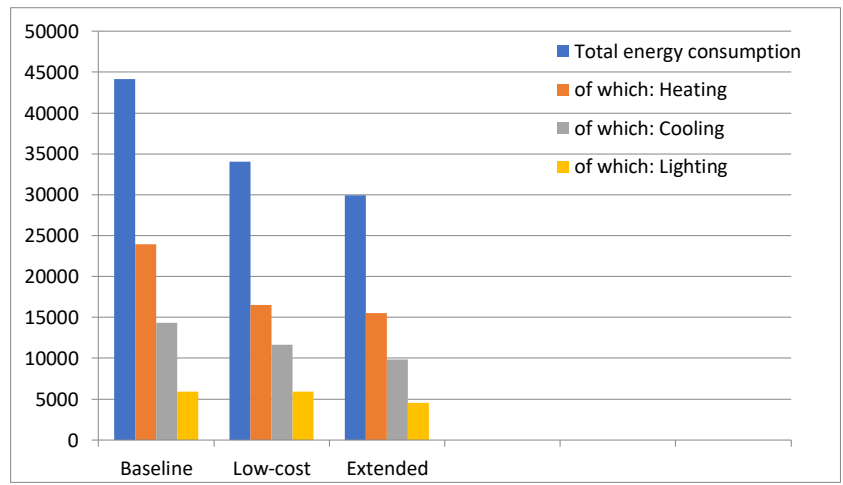

**Figure 3.** Annual energy consumption for heating, cooling and lighting; case of Stuttgart (kWh).

**Table 5.** Energy savings from home automation.

| City | Scenario | Energy Savings [kWh] | of Which: Heating [kWh] | of Which: Cooling [kWh] | of Which: Lighting [kWh] | Cumulated Energy Savings over 10-year-Period [MWh] |
|------|----------|---------|---------|---------|---------|---------|
| Algiers | Baseline | 0 | 0 | 0 | 0 | 0 |
| | Low-cost Installation | 6523 | 3281 | 3243 | 0 | 65 |
| | Extended Installation | 11,020 | 3539 | 6071 | 1410 | 110 |
| Stuttgart | Baseline | 0 | 0 | 0 | 0 | 0 |
| | Low-cost Installation | 10,092 | 7403 | 2689 | 0 | 100 |
| | Extended Installation | 14,222 | 8393 | 4466 | 1363 | 142 |

Comparing the saving results to the average annual consumption (e.g., for a German single-family-home, 30.2 MWh for average annual consumption without cooling [103]), it is clear that that the savings achieved already in the low-cost installation scenario amounts to over three years of saved energy consumption with increased comfort levels.

### 4.2. Economic Impacts

Not surprising, the energy saving impact directly translates into tangible economic impacts when applying indicators (3) payback period; (4) net present value and (5) internal rate of return. The results are summarized in Table 6. We present rounded values to underline that the figures only present the economic assessment of a simulation rather than real-life values.

**Table 6.** Summary of economic indicators.

| | Payback Period | | NPV (€) | | IRR (%) | |
|---|---|---|---|---|---|---|
| | Algiers | Stuttgart | Algiers | Stuttgart | Algiers | Stuttgart |
| Low-cost Installation | ~2 years, 4 months | ~2.5 months | 834 | 15,026 | 50 | 481 |
| Extended Installation | ~1 year,9 months | ~2.5 months | 1969 | 23,918 | 58 | 439 |

Given the low upfront investment costs, the energy savings achieved through home automation directly translate into highly profitable overall investments as presented by the net present value and the internal rate of return. This result implies that a homeowner who does not implement home automation is missing out on a highly beneficial investment. Note that the payback period in our calculations is highly sensitive to energy prices.

### 4.3. Social Impacts

Measuring the social dimension of energy savings proves more challenging than establishing the other indicators. As discussed in Section 3, we consider the additional disposable income, i.e., the annual monetary savings, as key indicator to track social impacts. Given the large amount of energy savings, the amount of additional disposable income is considerable in both cases, as can be seen in Figure 4.

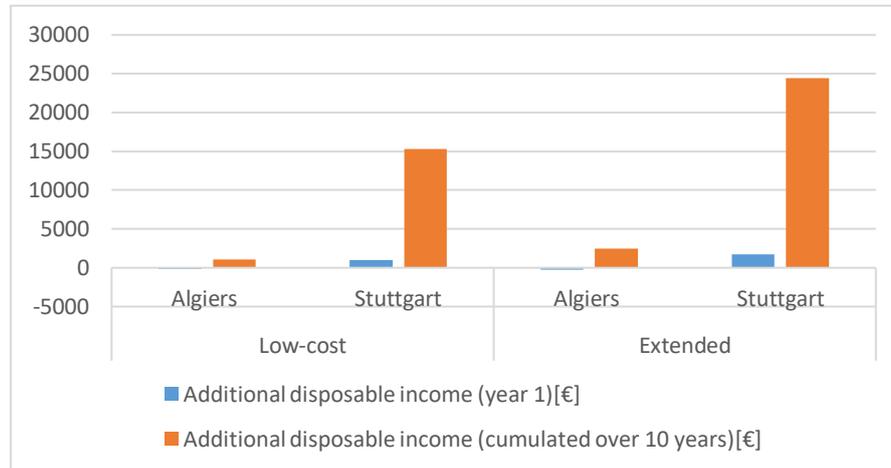

**Figure 4.** Additional disposable income.

The overall lifecycle analysis again highlights the social benefits of increased energy efficiency through smart home solutions, as discussed by Ürge-Vorsatz et al. and Kerr et al. [7,44,50]. However, when looking at the change of disposable income in year one, the Algiers case shows clearly the need to finance the additional investments. This mirrors the financing barrier of energy efficiency, which deters many households from engaging in increased energy efficiency solutions, as they are forced to live from a short-term perspective.

### 4.4. Environmental Impacts

Regarding the assessment of environmental impacts yielded by smart home solutions, we proposed indicator (7) in Section 3. This indicator applies environmental emission coefficients per several types of emissions to the achieved energy savings. By this, a rough estimate of emission and resource savings can be determined. The results are shown in Figures 5 and 6.

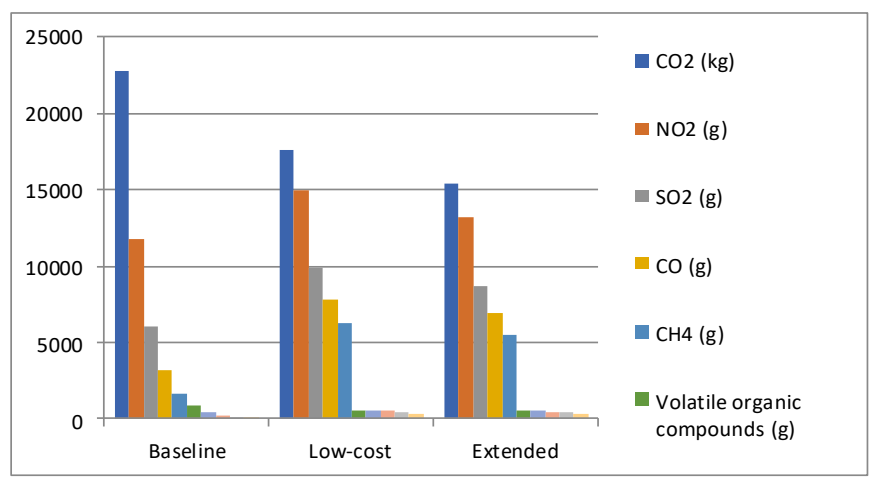

**Figure 5.** Environmental impacts - Algiers

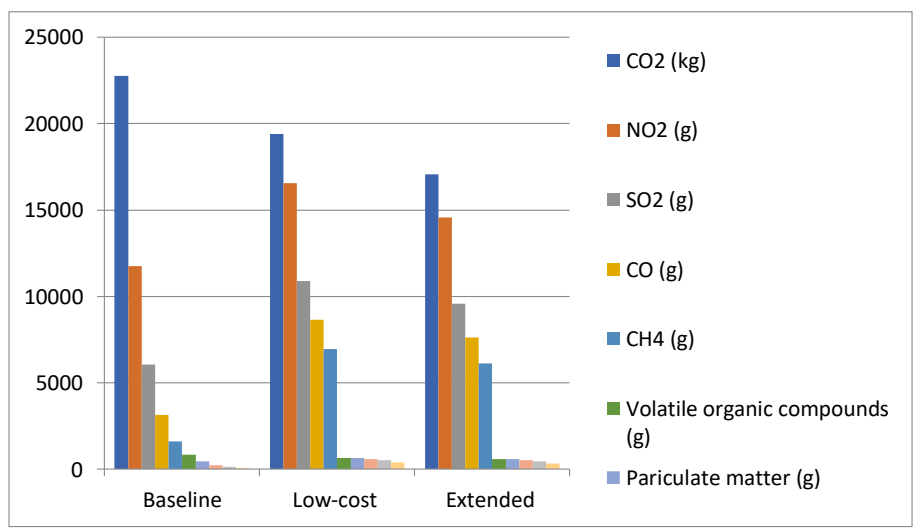

**Figure 6.** Environmental impacts—Stuttgart.

Beside the overall benefits of implementing energy efficiency solutions to the environment, the results above help to make tangible one's personal engagement in the welfare of the environment. The model household in our case could receive certification that by applying extended smart home solutions, it had contributed individually to saving $CO_2$ emissions by some 7.3 t $CO_{2eq.}$ (Algiers) and 5.7 t $CO_{2eq.}$ (Stuttgart). Similar considerations apply for other greenhouse gases and local air pollutants, which represent strong concerns for citizens and local authorities. This benefit is a further motive for taking up smart home solutions, thereby saving energy and minimizing the individual's footprint on the environment.

## 5. Discussion

We simulated the efficiency benefits provided by home automation over a 10-year life cycle of energy consumption from the lighting, heating and cooling of a model single-family home situated in two different climate regimes and two different regulatory environments—Germany and Algeria. Germany has a continental climate (Koeppen-Geiger classification Cfb), while Algeria has a Mediterranean climate (classification Csa). Energy prices paid by homeowners in Germany are set by the market, with network fees charged to finance the transition to sustainable energy; energy prices paid by homeowners in Algeria are supported by state subsidies. We were able to develop a set of indicators for efficiency benefits that can serve the interests of homeowners living under these two dissimilar conditions equally well. While our simulations do not permit a detailed comparison of benefits offered in the two countries, our findings show clearly that the return on investment in energy efficiency is highly sensitive to energy policy choices, such as energy taxation or standards for building refurbishment.

### 5.1. Homeowner Benefits Indicators

We have shown that indicators for economic benefits available to the homeowner can easily be derived by drawing on a set of standard indicators well established in cost-benefit analysis [104]. We have shown that applying specific emission conversion factors to energy savings allows a meaningful picture to emerge of the environmental impact of individual investments in efficiency. This is an approach that allows homeowners to assess the environment impact of their energy decisions at a scope and level of detail going well beyond the standard estimations of $CO_2$ savings currently made in both countries. Such detail serves homeowners' desires for more active engagement with the energy future of their families.

We have also shown that energy efficiency benefits can be positioned within the economic and environmental self-interests of a homeowner to make investments in efficiency decidedly attractive.

Our findings also show a positive return on investment for social benefits, which however is not as readily quantifiable with standard indicators.

The benefits across all categories are highly correlated, underlining again the necessity to broaden information campaigns to include a comprehensive assessment of the full spectrum of multiple benefits from energy efficiency. At present neither the Algerian nor the German government uses this opportunity to promote to their citizens the personal multiple benefits households can receive by improving home energy efficiency. Non-invasive promotion could be done by setting up online calculation tools that capture the full set of benefits accruable to the homeowner. Given that the indicators we applied are well established and easy to use, they could be included in existing online tools to further drive home the personal benefits available to the individual homeowner who takes steps toward more efficient use of energy.

Our findings are consistent with macroeconomic research that has shown how multiple-benefit analysis makes the case for energy efficiency even stronger [1,7,46,50,105-110]. This principle is well established in the literature at the macroeconomic level. Our findings show that multiple efficiency benefits are tangible and realizable for the individual homeowner as well. We find the use of extended smart home technology provides sizable benefits across all indicators, but even a low-cost installation can provide compelling economic benefits to the individual homeowner. The more individual homeowners can recognize and realize the multiple benefits available to their families through improved management of energy use, the easier it will be for countries to reach their long-term energy policy goals.

*5.2. Limitations and Further Research*

For several reasons, our findings can only serve as a first approximation of the role of multiple energy efficiency benefits at the homeowner level, but they clearly invite further elaboration by subsequent studies. Our analysis managed to identify and apply only a subset of relevant indicators from the benefit categories known in the literature. A need exists to broaden the indicators for social impacts so that benefits such as health improvements, home safety and improved occupant comfort can be captured. This is especially important given the pivotal role these aspects play in household buying decisions of smart home systems [111].

We also chose to simulate only one home configuration, namely the two-level single-family home with two bedrooms and one bath, representing a common starter home, which for a good number of families remains their home for life. We modelled this home using a set of standardized template settings. While this does represent the dwelling type with the largest energy saving potential in both countries, our work is best taken as a starting point for extended investigation into other dwelling types. The positive effects of the multiple benefits of energy efficiency have persuasive potential for all property owners.

Our simulations modelled only two climate regimes and two regulatory environments. Subsequent research could perform sensitivity analyses to evaluate the impact on our indicators of variations in energy prices, discount rates and payback expectations from homeowners.

Our simulations also did not monetize environmental impacts, as our set of benefit indicators could only capture the impact on air pollutants. Even without monetization, however, our findings demonstrate for homeowners the improvements they can make in their environmental footprint by improving the energy efficiency of their home.

## 6. Conclusions and Recommendations

We simulated the multiple benefits of energy efficiency brought forward by smart home solutions. The predominantly macroeconomic research orientation in the literature abstracts away homeowner concerns into aggregate probabilities. We chose instead to study impacts on the individual homeowner, who makes the decision to invest in energy efficiency [112]. From established lines of multiple benefit analysis, we derived impact indicators for the homeowner, which we then analyzed over a 10-year lifetime of energy use by a model single-family home in the climate zones and regulatory environments of northern Algeria and southern Germany. We compiled well-established, easy-to-use benefit-indicators to assess the economic, environmental and social benefits of smart home solutions. We find monetizable benefits for the homeowner in resource use, investment opportunity, and social impact, as well as quantifiable benefits in climate change mitigation.

Further research recommended by our study would investigate in greater detail both social and environmental impact indicators. The quality of an indicator depends on how comprehensively it captures impacts in the relevant benefit category. Social impact is a particularly challenging metric to develop, which suggests a fruitful line of future research.

Policy recommendations also follow from our findings. The story of monetizable benefits for homeowners who improve their home energy efficiency using smart technology is one that needs to be told. It is fair to say that policy goals will move that much closer to the self-interests of individual homeowners, the better those homeowners understand the multiple benefits they gain from energy efficiency. In particular, the regulatory process of rolling out smart meters and defining smart home standards can be used for this in several ways:

(1) Upgrade existing online calculation tools for energy efficiency to display the full set of multiple benefits;
(2) Accompany the present roll-out of smart meters with standards and field tests so metering and sensor data can be used to quantify additional benefits;
(3) Align government support schemes to promote the realization of multiple benefits, such as "first aid" home automation investment support for energy poor households with least cost government investment.

The current simulation tools that cover energy and $CO_2$ emissions, or economic returns on efficiency investments can be enlarged to cover the full set of multiple benefits. With the emergence of big data from smart meters and local sensors, our simulation approach can be expanded not only to cover a wide range of multiple benefits, but also to improve a homeowner's ability to perform comprehensive what-if analyses when evaluating investment decisions.

**Author Contributions**: Conceptualization, M.R., D.D. and R.L.; Methodology, R.L., M.R.; Software, D.D., R.L.; Software validation, D.D., R.L.; Formal analysis, R.L., M.R. and D.D.; Data curation, R.L.; Writing—original draft, review and editing, M.R., R.L. and D.D.; Visualization, R.L.; Funding acquisition, D.D. and M.R

**Funding:** The present contribution was developed within the framework of the Arab German Young Academy of Sciences and Humanities (AGYA), partly in Nuertingen Geislingen University, Germany, and CERSIT research Centre, Algeria. It draws on financial support of the German Federal Ministry of Education and Research (BMBF) grant 01DL16002, as well as the Algerian Ministry of Higher Education and Research. The authors are very grateful for this support.

**Conflicts of Interest:** The authors declare no conflict of interest.